\documentclass{emulateapj}
\usepackage{psfig}
\usepackage{footmisc}

\newcommand{\ssst}{\scriptscriptstyle}
\newcommand{\E}[1]{\times 10^{#1}}
\newcommand{\etal}{et al.}
       
\newcommand{\RA}[3]{{#1}^{{\rm h}}{#2}^{{\rm m}}{#3}^{{\rm s}}}
\newcommand{\decl}[3]{{#1}^{\circ}{#2}'{#3}''}

      \newcommand{\ps}{\,{\rm s}^{-1}}
\newcommand{\yr}{\,{\rm yr}}    \newcommand{\Msun}{M_{\odot}}
\newcommand{\cm}{\,{\rm cm}}    \newcommand{\km}{\,{\rm km}}
\newcommand{\kms}{$\km\ps$}
\newcommand{\kpc}{\,{\rm kpc}} 
\newcommand{\erg}{\,{\rm erg}}        \newcommand{\K}{\,{\rm K}}
    \newcommand{\keV}{\,{\rm keV}}

\newcommand{\um}{\,\mu\rm m}

\newcommand{\nel}{n_{e}}        \newcommand{\NH}{N_{\ssst\rm H}}
\newcommand{\no}{n_{\ssst 0}}

\newcommand{\rs}{r_{s}}         \newcommand{\vs}{v_{s}}
\newcommand{\nH}{n_{\ssst\rm H}}        \newcommand{\mH}{m_{\ssst\rm H}}
 \newcommand{\NHH}{N({\rm H}_{2})}
\newcommand{\VLSR}{V_{\ssst\rm LSR}}
       
\newcommand{\ROSAT}{{\sl ROSAT}} \newcommand{\Spitzer}{{\sl Spitzer}}

\newcommand{\XMM}{{\sl XMM}}
\newcommand{\Newton}{{\sl Newton}}
\newcommand{\du}{d_{5}} 
    
\newcommand{\Rb}{R_{\rm b}}

\newcommand{\Ha}{H$\alpha$}
\newcommand{\snr}{Kes~78}
\newcommand{\twCO}{$^{12}$CO}   \newcommand{\thCO}{$^{13}$CO}
\newcommand{\Jotz}{$J$=1--0}    \newcommand{\Jtto}{$J$=2--1}

\shorttitle{}

\begin{document}

\title{MOLECULAR ENVIRONMENT AND AN X-RAY SPECTROSCOPY OF
    SUPERNOVA REMNANT KESTEVEN~78 }

\author{
 Ping Zhou\altaffilmark{1}
 and Yang Chen\altaffilmark{1,2,3}
}
\altaffiltext{1}{Department of Astronomy, Nanjing University, Nanjing~210093,
       China}
\altaffiltext{2}{Key Laboratory of Modern Astronomy and Astrophysics,
 Nanjing University, Ministry of Education, China}
\altaffiltext{3}{Author to whom any correspondence should be addressed.}

\begin{abstract}


We investigate the molecular environment of the Galactic
supernova remnant (SNR) Kesteven~78
and perform an \XMM-\Newton\ X-ray spectroscopic study
for the northeastern edge of the remnant.
SNR~\snr\ is found to interact with the molecular clouds (MCs)
at a systemic local standard of rest velocity of $81\km\ps$.
At around this velocity, the SNR appears to contact a long
molecular strip in the northeast and a large cloud
in the east as revealed in the \thCO\ line,
which may be responsible for the radio brightness
peak and the OH maser, respectively.
The \twCO-line bright region morphologically matches the eastern
bright radio shell in general,
and the SNR is consistent in extent with a CO cavity.
Broadened \twCO\ line profiles discerned in the eastern maser region
and the western clumpy molecular arc
and the elevated \twCO~\Jtto/\Jotz\ ratios along the SNR boundary
may be signatures of shock perturbation in the molecular gas.
The SNR-MC association places the SNR at a kinematic distance of 4.8~kpc.
The X-rays arising from the northeastern radio shell are
emitted by underionized hot ($\sim1.5\keV$), low-density
($\sim0.1\cm^{-3}$) plasma with solar abundance,
and the plasma may be of intercloud origin.
The age of the remnant is inferred to be about 6~kyr.
The size of the molecular cavity in \snr\ implies an initial
mass around $22\Msun$ for the progenitor.
\end{abstract}


\section{Introduction} \label{sec:intro}
Born in giant molecular clouds (MCs), massive stars evolve from
collapsing molecular cores to core-collapse supernovae.
During their short lifetimes, they evacuate a cavity with their energetic
winds and ionizing radiation, and later the supernova shocks will
often collide with the cavity wall.
In the past three decades, dozens of supernova remnants (SNRs) were found to be in physical
contact with MCs (see Jiang et al.\ 2010 and references therein).
In particular, the 1720 MHz OH masers, collisionally excited by the
shock-heated molecular gas, have been regarded as the signposts of
SNR-MC interaction (e.g., Frail \etal~1996; Lockett \etal~1999;
Frail \& Mitchell 1998; Wardle \& Yusef-Zadeh 2002).
Recently, several GeV and TeV sources were found to spatially
correspond to MCs that are associated with SNRs, such as IC443 and W28, in
which $p-p$ collision plays an important role in the gamma-ray emission.
Detailed investigation of the molecular environment of the interacting SNRs
is therefore of great interests.

Kesteven~78 (G32.8--0.1), with a partly brightened radio shell,
was identified as a Galactic SNR by Caswell et al.\ (1975)
based on the 408 and 5000MHz observations.
A distance of $\sim8$~pc was estimated using the radio surface brightness
-- diameter ($\Sigma$--$D$) relationship (Caswell et al.\ 1983).
A 21cm HI observation with a resolution $2'\times130'\times6.3\km\ps$
was made by Gosachinskii \& Khersonskii (1985), which suggested a
corresponding local standard of rest (LSR) velocity of $90\km\ps$,
a distance of 9~kpc, and an age of 12~kyr for the SNR.
Kassim (1992) derived the spectral index as $-0.5$.
A single 1720~MHz hydroxyl (OH) maser spot at LSR velocity $86.1\km\ps$
was detected on the radio shell by Koralesky et al.\ (1998) in the
Very Large Array (VLA) observations, which indicates a shock interaction
with MCs;
a distance of 5.5 or 8.8~kpc was then suggested on the assumption
that the shock propagation is perpendicular to line of sight (LOS).
A classical antisystematic S-shaped profile of Stokes~V  was also
detected, suggesting the presence of Zeeman splitting.
A millimeter-wavelength \twCO\ (\Jotz) observation was carried out by
Zhou et al.\ (2007), which suggests that an eastern molecular gas
detected in the velocity range 72--$88\km\ps$ is related to the SNR,
and no line broadening was found from it.
In a \Spitzer-IRAC mid-infrared (IR) survey of Galactic SNRs,
which was made to search for IR counterparts with existing
radio images, \snr\ is classified as ``not detected but confused" one
(Reach et al.\ 2006).
In the optical band, filamentary and diffuse emission is detected in
the SNR; the [SII]/\Ha\ ratio higher than 1.2 is extensively
found, suggesting that the optical emission arises from the shock-heated gas
(Boumis et al.\ 2009, B09).
An extended very high energy (VHE) $\gamma$-ray source HESS J1852$-$000 is
revealed by the H.E.S.S team to be close to the east edge of the
remnant.\footnote{http://www.mpi-hd.mpg.de/hfm/HESS/pages/home/som/2011/02/
\label{fn:hess}}

Motivated by the indication of the OH maser for SNR--MC interaction;
a hint of the previous CO observation for a related MC in the east;
and the irregular morphology of the SNR, which implies a complex,
non-uniform ambient medium, we examine the detailed distribution of
the environmental molecular gas and its physical relation with \snr,
based on both our new and the archival observations of
three CO rotational transitions and
the archival data of the \XMM-\Newton\ X-ray observation.
In Section~2, we briefly describe the observations and data reduction.
Our results are presented in Section~3. Physical properties of the SNR
are discussed in Section~4. The conclusions are summarized in Section~5.

\section{Observations and Data Reduction}
\subsection{CO Observations and Data}
The observations of millimeter molecular emissions toward SNR~\snr\
were first made in 2009 November with the 13.7 m millimeter-wavelength
telescope of the Purple Mountain Observatory at Delingha (hereafter
PMOD), China. A superconductor-insulator-superconductor (SIS) receiver
and two acousto-optical spectrometers (AOSs) were used to simultaneously
observe the \twCO\ (\Jotz) line (at 115.271~GHz) and the \thCO\ (\Jotz)
line (at 110.201~GHz).
We mapped the radio bright shell region of \snr\ with $1'$ grid spacing and a $39'\times27'$ large region
centered at ($\RA{18}{51}{03}.97$,
$\decl{-00}{12}{22}.5$, J2000.0) with $2'$ grid spacing.
The AOS bandwidth and velocity resolution are 145~MHz and $0.37\km\ps$
for \twCO (\Jotz) and 43~MHz and $0.11\km\ps$ for \thCO\ (\Jotz).
During the observation epoch, the half-power beam width (HPBW)
of the antenna was $56''$,
the main beam efficiency was about 62\% at zenith,
and the pointing accuracy was better than $5''$.
The typical system temperature was around 230~K.
The observed LSR velocity ranges were
$-102$ to $276\km\ps$ for \twCO\ (\Jotz) and 28 to $143\km\ps$ for \thCO\
(\Jotz).

The follow-up observation was made in the \twCO\ (\Jtto) line
(at 230.538~GHz) during 2010 January and February
using the K\"{o}lner Observatory for Submillimeter Astronomy (KOSMA)
3m submillimeter telescope in Switzerland.
An SIS receiver and a medium-resolution AOS spectrometer were used.
A $31'\times31'$ area centered at ($\RA{18}{51}{03}.97$,
$\decl{-00}{08}{22}.5$, J2000.0) was mapped with a grid spacing
of $1'$ in on-the-fly mode.
The HPBW of the telescope was $130''$, the main beam efficiency was
54\% during our observation, and the pointing accuracy was about $10''$.
The AOS bandwidth and velocity resolution were about
300~MHz and $0.2\km\ps$.

All the CO data were reduced with GILDAS/CLASS package
\footnote{http://www.iram.fr/IRAMFR/GILDAS} and analyzed with IDL and
KARMA (Gooch, 1996).
After the baseline subtraction and the calibration for main
 beam efficiency and elevation,
the spectra were resampled to a uniform velocity resolution of $0.5\km\ps$
and the data were convolved to a uniform beam size of $2'$.
Then the mean rms noise levels of the main beam brightness temperature were
0.18, 0.20, and 0.30K for the \twCO\ (\Jotz), \thCO\ (\Jotz), and
\twCO\ (\Jtto) lines, respectively.

\thCO~(\Jotz) emission data from the Boston University--Five College Radio
Astronomy Galactic Ring Survey (BU-FCRAO~GRS; Jackson et al.\ 2006)
were also used. They offer a $46''$ angular resolution on a $22''$ grid
and a 0.21~\kms velocity resolution.

\subsection{Data of X-Ray and Other Wavebands}
The \ROSAT\ X-ray image of the SNR \snr\ region was gained from
the \ROSAT\ All Sky Survey broadband data.
The \XMM-\Newton\ data of partial region of \snr\ were obtained from
an observation toward the Galactic ridge centered at
($\RA{18}{51}{44}.67$, $\decl{00}{08}{58}.5$, J2000),
which was carried out on 2003 November 21--22 (ObsID: 0017740501,
PI: F.\ Azita Valinia) in the full frame mode and with the medium filter.
The EPIC data were reprocessed with the Science Analysis System
software (SAS, version 9.0.0).
After removing 
time intervals with heavy proton flarings,
net exposures of 29~ks and 27~ks remained for EPIC-MOS and EPIC-PN,
respectively, and were used for analysis.
The \XMM-\Newton\ field of view (FOV) covers only the northeastern
boundary region of \snr, which is the brightest region in the 1.4GHz radio
continuum emission.

The \Spitzer\ $24\um$ mid-IR observation used here was carried out
as 24 Micron Survey of the Inner Galactic Disk Program
(PID: 20597; PI: S. Carey) with the Multiband Imaging Photometer.
The $24\um$ Post Basic Calibrated Data
were obtained directly from the \Spitzer\ archive.
The 1.4 GHz radio continuum emission and the HI-line data
were obtained from the archival VLA Galactic Plane Survey
(VGPS; Stil et al.\ 2006).

\section{Results}
\subsection{Molecular Environment}
\subsubsection{The OH Maser Point and MCs at $\VLSR\sim81\km\ps$}
  \label{S:vlsr}

\begin{center}
\begin{figure}[tbh!]
\centerline{ {\hfil\hfil
\psfig{figure=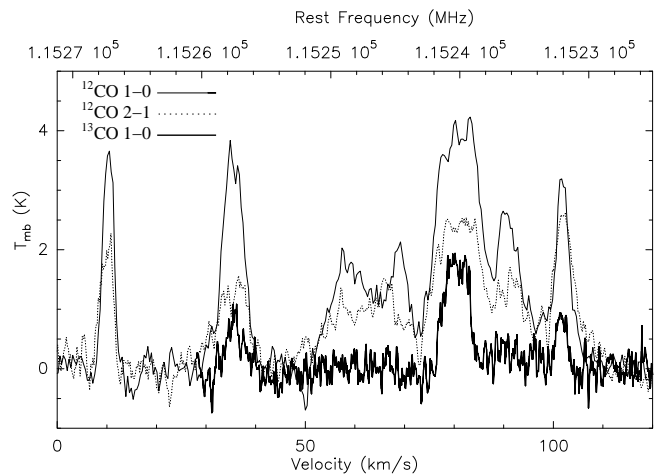,height=2.5in,angle=0, clip=}
\hfil\hfil}}
\caption{
\twCO~(\Jotz) (solid line),\twCO~(\Jtto) (dotted line),
and \thCO~(\Jotz)(thick solid line) spectra at the maser
point. The upper $x$-axis refers to the rest frequency of
the \twCO~(\Jotz) line. All the spectra are not convolved to
a uniform beam size.
}
\label{f:maser_spec}
\end{figure}
\end{center}

\begin{center}
\begin{figure*}[tbh!]
\centerline{ {\hfil\hfil
\psfig{figure=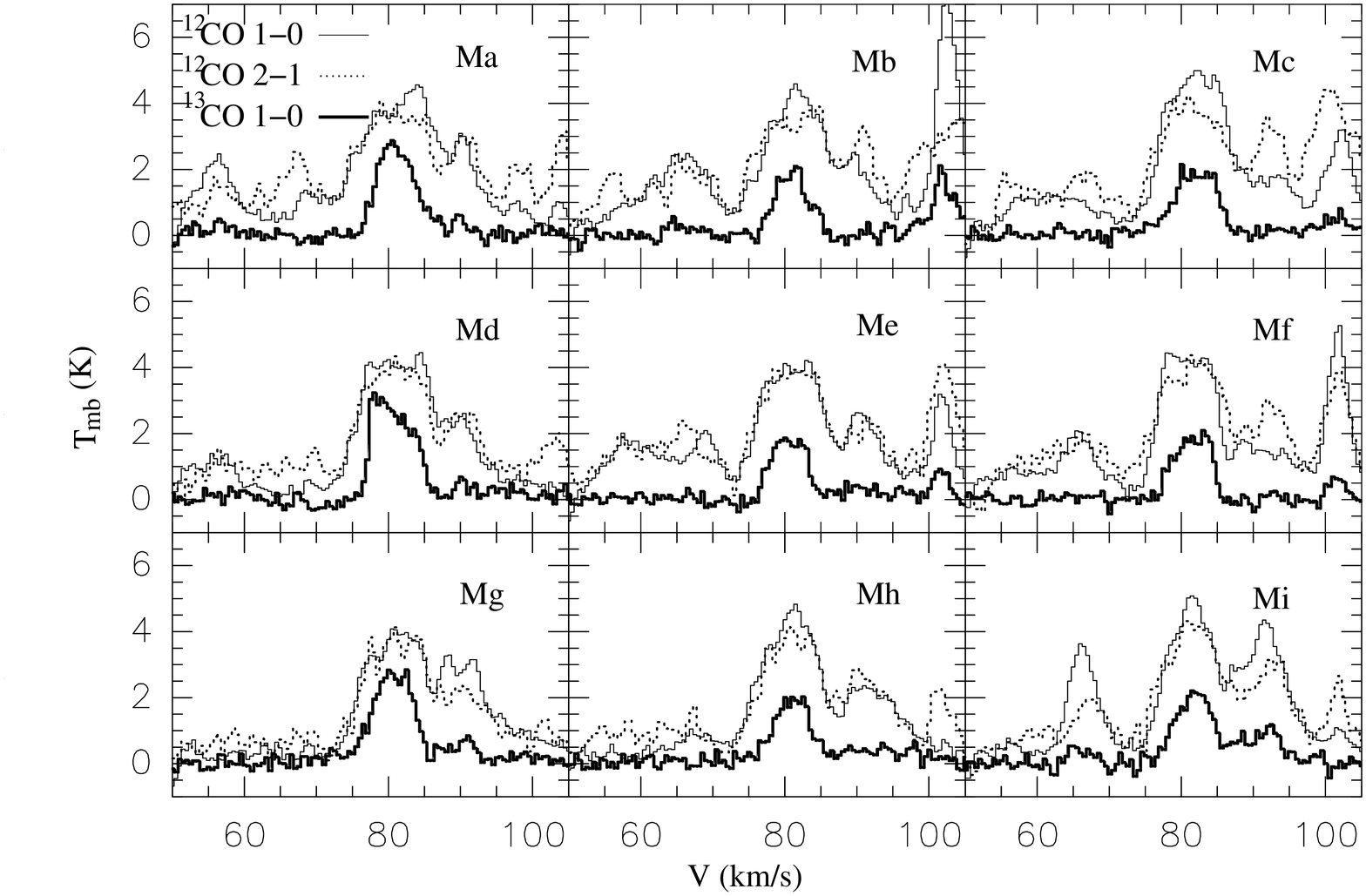,height=3.in,angle=0, clip=}
\hfil\hfil}}
\caption{Grid of CO spectra
restricted to the velocity range $50\km\ps$ to $105\km\ps$
in the 1720~MHz OH maser region with $1'$ spacing
(labeled as the box ``M'' in the bottom right panel of Figure~\ref{f:COmap}). Note that the
\twCO\ (\Jtto) lines have been multiplied by a factor of 1.6 so as to scale
the $\sim81\km\ps$ \twCO\ (\Jtto) peak of the maser point (``Me")
with the \twCO\ (\Jotz) peak. All the spectra are not convolved to
a uniform beam size.
}
\label{f:maser_grid}
\end{figure*}
\end{center}

We first focus on the molecular gas indicated by the 86 \kms 1720~MHz
OH maser located at ($\RA{18}{51}{48}.04,\decl{-00}{10}{35}$).
The CO spectra at the maser point are given in Figure~\ref{f:maser_spec}.
There are four prominent peaks at $\VLSR\sim10$, 37, 81, and $102\km\ps$.
The LSR velocity of the maser, $86\km\ps$, is in the red (right) wings of
the $81\km\ps$ \twCO\ lines, in which \thCO\ emission is insignificant.
A few secondary \twCO\ components ranging from $52\km\ps$ to
$96\km\ps$, with few \thCO\ counterparts, seem to be linked
with the prominent component peaked at $81\km\ps$.
As \thCO\ emission, usually optically thin, arises in the quiescent
dense gas along the whole LOS,
the lack of significant \thCO\ features at the corresponding LSR
velocities indicates that these \twCO\ secondaries are likely to
represent the disturbed gas deviating from the systemic velocity
$\sim81\km\ps$, as complicated broadened wings.
Similar CO line profiles are also present at points
close to the OH maser (point ``Me"),
as seen in the grid (Figure~\ref{f:maser_grid}).
By scaling the $\sim81\km\ps$ \twCO\ (\Jtto) peaks to match the \twCO\ (\Jotz)
peaks, with the maser (``Me") as the fiducial point, relative enhancement
of the \twCO\ (\Jtto) is seen at some points
(e.g., ``Mb", ``Mc", and ``Mf" at $\sim90\km\ps$
and ``Ma", ``Mc", and ``Me" at $\sim67\km\ps$),
which may be indicative of relatively warm shocked gas.
The shock interaction is also favored by the elevated
\twCO~\Jtto/\Jotz\ ratio
at 64--$68\km\ps$ in the maser vicinity (see below).

\begin{center}
\begin{figure*}[tbh!]
\centerline{ {\hfil\hfil
\psfig{figure=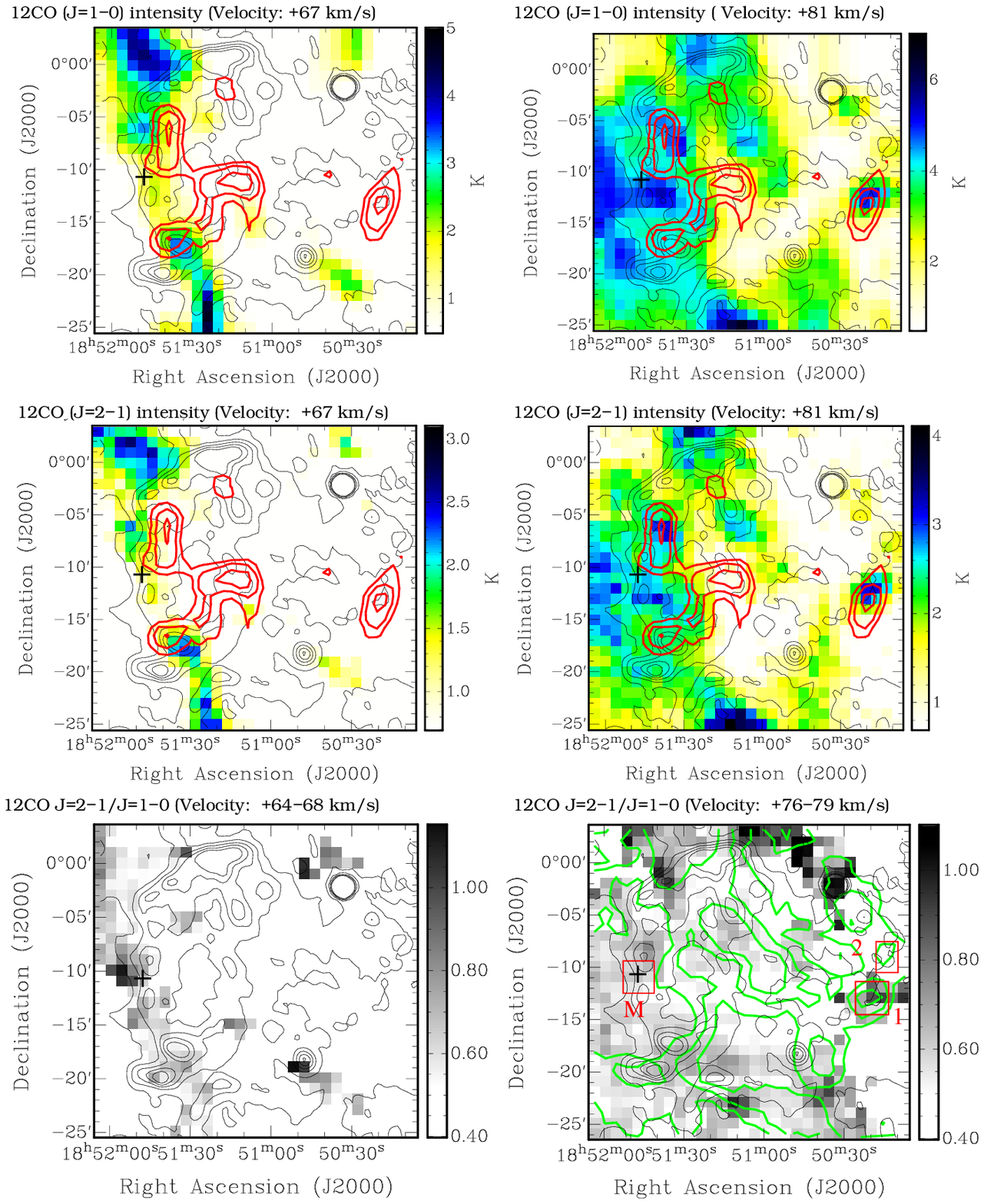,height=7.2in,angle=0, clip=}
\hfil\hfil}}
\caption{
Upper and middle rows:
intensity maps of \twCO~(\Jotz) and \twCO~(\Jtto)
 at the $\VLSR\sim67\km\ps$ and $\VLSR\sim81\km\ps$.
Thick contours (red) represent the \ROSAT\ 0.1--2 keV X-ray
emission with levels of 0.22, 0.26, and 0.30 counts~pixel$^{-1}$
(after being smoothed with a $3'$ Guassian kernel).
Bottom row: \twCO \Jtto/\Jotz\ ratio maps in 64--$68\km\ps$ and
76--$79\km\ps$, after convolving the PMOD \twCO~(\Jotz) to the same beam size
as the KOSMA \twCO~(\Jtto) data. The thick contours (green) represent
the \twCO~(\Jtto) emission at 80--$84\km\ps$ interval with levels
6.5, 10.1, and 13.7 K \kms. Here both \twCO (\Jtto) and \twCO (\Jotz)
data used achieve a signal-to-noise ratio $>3$.
Thin contours (black) of VGPS 1.4GHz continuum emission are overlaid
in six panels with  levels of 16, 20, 24, 28 and 32K.
The plus sign in each panel denotes the OH maser point.
The boxes, labeled with ``M", ``1'', and ``2", define the OH maser region,
the cores of ``clump 1", and ``clump 2", the spectra of which are shown in
Figures~\ref{f:maser_grid}, \ref{f:clump1}, and \ref{f:clump2}.
}
\label{f:COmap}
\end{figure*}
\end{center}

By inspecting the intensity maps in the entire velocity range, however,
we only found two velocity intervals (around $67\km\ps$ and $81\km\ps$)
in which the \twCO\ emission demonstrates more or less morphological
correspondence with the SNR.
As seen in Figure~\ref{f:COmap},
at round 81 \kms, the CO emission exhibits a cavity structure,
which is consistent with the SNR as delineated by the radio contours.
The eastern bright radio shell of the SNR overlaps with
most of the region of strong \twCO\ emission,
which covers the maser point.
The radio emission fades from east to west,
and a \twCO\ arc composed of a few clumps is located along the
faint radio boundary in the west. In the SNR boundary,
four prominent \ROSAT\ 0.1--2.4~keV X-ray patches are seen,
two of which are located along the western radio shell and
one of which is coincident with one of the western \twCO\ clumps
(``clump 1", see Section \ref{S:arc}).

At $\VLSR\sim67\km\ps$ (Figure~\ref{f:COmap}), a string-like
\twCO\ feature extends from the northeast to the south
in the FOV, also covering the maser point,
and most of it seems to follow the western radio shell.
However, two bright ends of the string are located outside the
remnant, and little \twCO\ emission corresponds to the radio brightness
peak on the northern shell.

The integrated CO-line intensity ratio between the different transitions
is used as kinematic evidence of interaction between
SNR shocks and MCs (e.g., Seta \etal\ 1998; Jiang \etal\ 2010).
Elevated \twCO~\Jtto/\Jotz\ line ratios $\sim1$ are also present
on the boundary in the two velocity intervals of concern.
At 76--79 \kms, the \twCO~\Jtto/\Jotz\ ratios are apparently
elevated to $\ga1$ (compared with $\sim0.4$--0.6 in average)
at several positions along the radio boundary.
Apart from the northwestern and southern boundaries,
where an incomplete grid
of points spaced every $2'$ was observed in PMOD for \twCO~(\Jotz) and
thus the ratios there should be taken with caution, the ratio
elevation in the northeast and the west is noteworthy.
The northeastern ratio elevation ($\approx1$) at
around ($\RA{18}{51}{37}$, $\decl{-00}{01}{22}$, J2000) is located in a bright patch
of radio emission, and the western elevation ($\ge0.9$) around
($\RA{18}{50}{16}$, $\decl{-00}{13}{00}$, J2000) is coincident
with a molecular clump (``clump 1", see Section \ref{S:arc}).
In some interacting SNRs, the line ratios are observed to exceed unity
(see Dubner \etal~2004), as found in SNR~3C396, however, due to some reasons
such as beam dilution of the small filling factor of the \twCO~(\Jtto)
emitting region compared to that for \twCO~(\Jotz),
the observed relatively elevated ratios $>0.9$ can also be a good probe of the
SNR--MC interaction (Su \etal~2011).
At 64--$68\km\ps$, the line ratio is enhanced to $>1$ in a region
around ($\RA{18}{51}{55}$, $\decl{-00}{10}{00}$, J2000) adjacent to the OH maser point.
By comparison, the line ratios are low in the two bright ends, which are outside
the SNR boundary, of the eastern $67\km\ps$ \twCO\ string.
The high line ratio near the maser point is another signature of disturbance
of the molecular gas there by the SNR shocks and the gas
emission at around $67\km\ps$ is in the blue wings of the main-body clouds
at $81\km\ps$.

\begin{center}
\begin{deluxetable}{l|ccccc}
\tabletypesize{\footnotesize}
\tablecaption{Some Parameters of the Molecular Gas Associated with \snr}
\tablewidth{0pt}
\tablehead{
\colhead{Regions}  &\multicolumn{2}{c}{$N(\rm H_2) (10^{21} \cm^{-2})$}  & & \multicolumn{2}{c}{$M (\Msun)$} \\ \cline{2-3} \cline{5-6}
& X & LTE & & X & LTE
}
\startdata
Maser point & 8.8 & 8.5 &   \\
Clump 1 & 4.3 & 3.0 & & $1.9\E{3}~d_{5}^{2}$ & $1.3\E{3}~d_{5}^{2}$ 
%
\tablecomments{
Here two methods (indicated as ``X" and ``LTE", respectively) are used to estimate the
parameters of the molecular gas; see the last paragraph in Section~\ref{S:vlsr}.
}
\enddata

\label{T:mass}
\end{deluxetable}
\end{center}

\begin{figure}[tbh!]
\centerline{ {\hfil\hfil
\psfig{figure=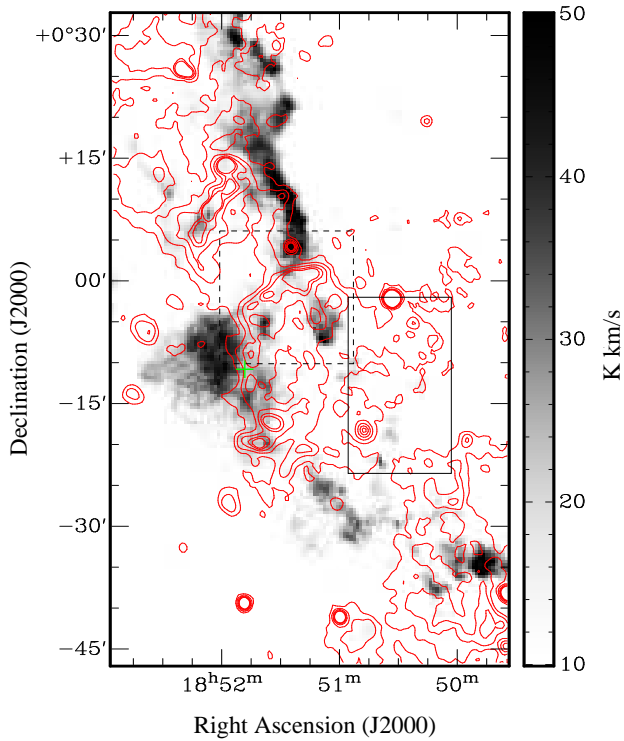,height=4.in,angle=0, clip=}
\hfil\hfil}}
\caption{
Molecular environment of \snr\ in the 80--$84\km\ps$ interval
(the BU-FCRAO~GRS \thCO~(\Jotz) observation), overlaid with
1.4~GHz radio contours at levels 16, 19.75, 23.5, 27.25, and 31~K.
The plus sign denotes the OH maser point. The two boxes indicate
the spatial ranges of the two maps given in Figure~\ref{f:closeup}.
}
\label{f:environ}
\end{figure}

\begin{figure*}[tbh!]
\centerline{ {\hfil\hfil
\psfig{figure=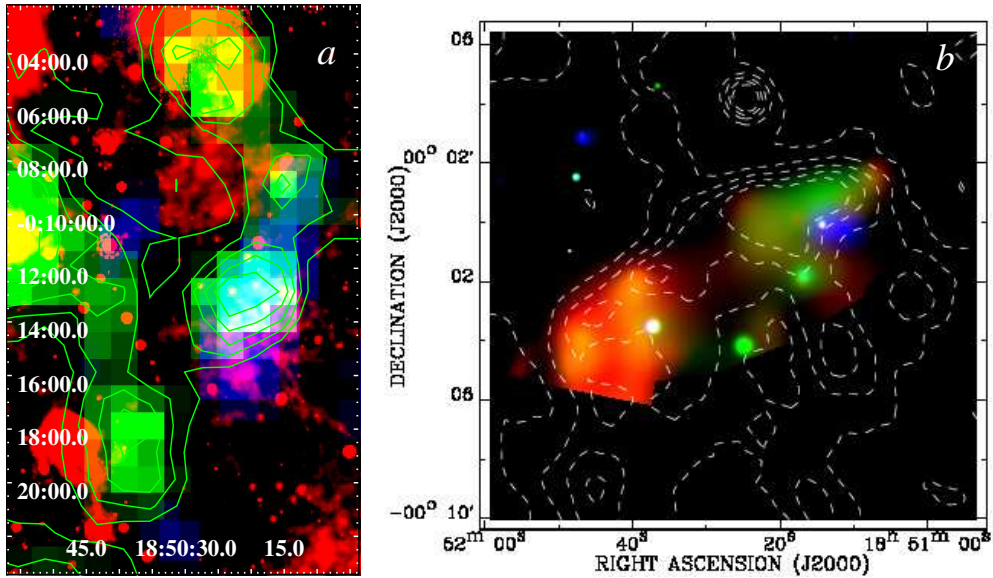,height=3.in,angle=0, clip=}
\hfil\hfil}}
\caption{
(a) A close-up tri-color image of the western molecular arc
(the region of the image is indicated by the solid box in
Figure~\ref{f:environ}). Red: \Spitzer\
$24~\um$ emission; green: intensity map of \twCO~(\Jtto) integrated from
80~\kms\ to 84\kms (overlaid with contours at levels of 6.5, 8.3, 10.1,
11.9, and 13.7~K \kms); and blue: \ROSAT\ 0.1--2.0~keV X-ray map
(smoothed to $3'$).
(b) Tri-color \XMM\ MOS X-ray image of the northeastern part of \snr\
(the region of the image is indicated by the dashed box in Figure~\ref{f:environ}).
The X-ray intensities in the 0.5--1.0, 1.0--2.0, and 2.0--8.0~keV bands
are coded in red, green, and blue, respectively.
The 1.4~GHz radio contours are at the same levels as those in
Figure~\ref{f:COmap}.
}
\label{f:closeup}
\end{figure*}

\begin{center}
\begin{figure*}[tbh!]
\centerline{ {\hfil\hfil
\psfig{figure=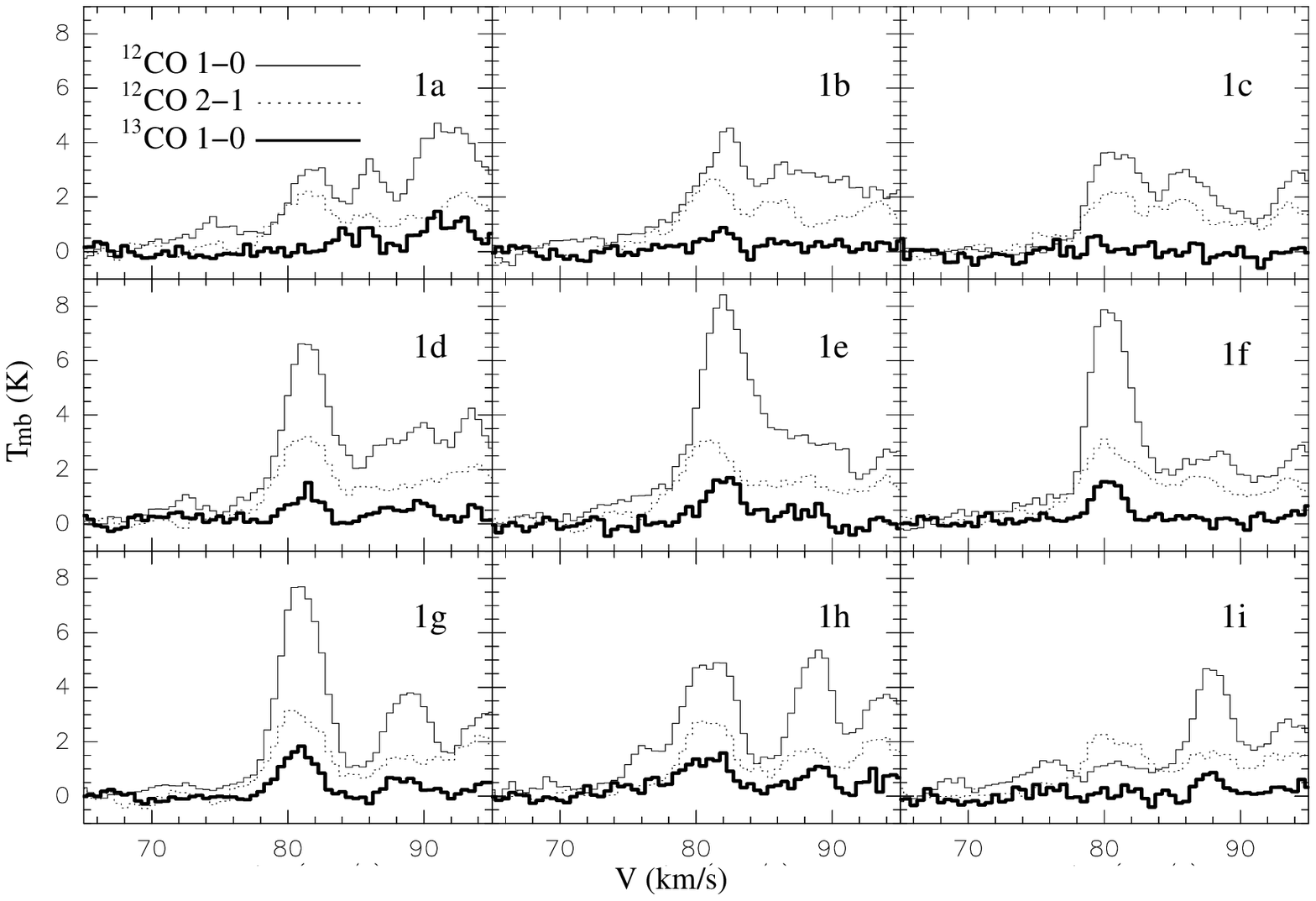,height=3.in,angle=0, clip=}
\hfil\hfil}}
\caption{
Grid of CO spectra
restricted to the velocity range $65\km\ps$ to $95\km\ps$
in ``clump 1'' with $1'$ spacing
(as labeled in the bottom right panel of Figure~\ref{f:COmap}).
All the spectra are not convolved to a uniform beam size.
}
\label{f:clump1}
\end{figure*}
\end{center}

\begin{center}
\begin{figure}[tbh!]
\centerline{ {\hfil\hfil
\psfig{figure=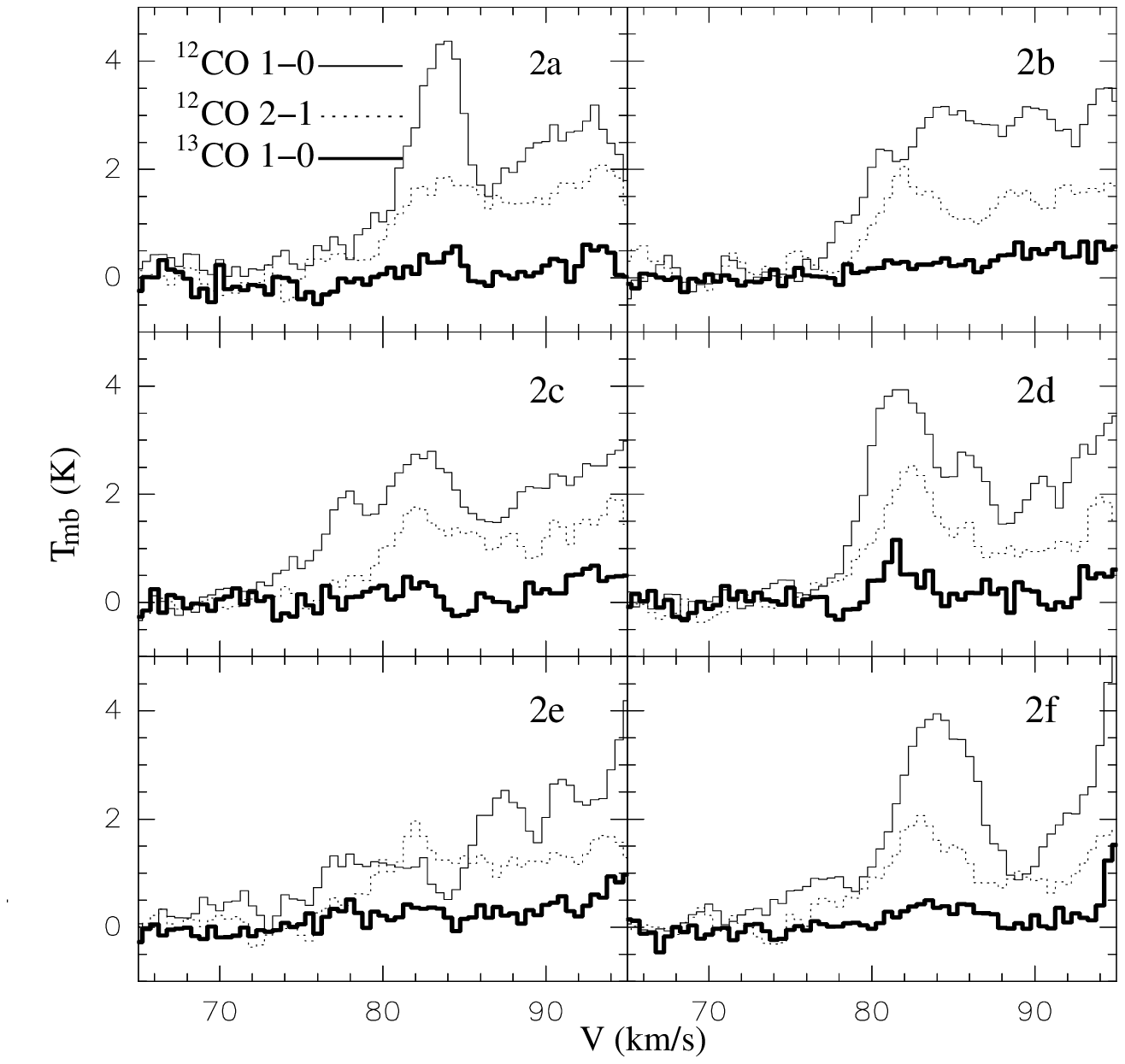,height=3.in,angle=0, clip=}
\hfil\hfil}}
\caption{
Grid of CO spectra
restricted to the velocity range $65\km\ps$ to $95\km\ps$
in ``clump 2'' with $1'$ spacing
(as labeled in the bottom right panel of Figure~\ref{f:COmap}).
All the spectra are not convolved to a uniform beam size.
}
\label{f:clump2}
\end{figure}
\end{center}

Figure~\ref{f:environ} shows a large FOV of the
\thCO\ (\Jotz) environment at 80--$84\km\ps$ around SNR \snr.
The SNR is seen to be located in the side of some MCs.
A long molecular strip in the northeast seems to connect the radio
shell, although the LOS projection effect
cannot be excluded, and the radio brightness peak along the northwestern
SNR boundary is located at the connection point.
The SNR also seems to be in contact with a molecular patch in the
east and the OH maser is located at the interface. The eastern
molecular patch seems to spatially correspond to some extent to
the extended VHE $\gamma$-ray source HESS J1852-000 (considering
the relatively large point-spread function of the H.E.S.S.
observation).\footref{fn:hess}
The \thCO\ emission is much fainter in the west of the remnant than in
the east.
This, together with the faint radio emission in the west of the
remnant, is consistent with a scenario that the remnant blows
out to a low-density region.
We also inspected the \thCO\ (\Jotz) emission at around
$67\km\ps$, but no distinct molecular structure was seen.


The above evidence from CO emission, such as the morphological correspondence
of the CO emission with the SNR, the enhanced \twCO\ \Jtto/\Jotz\ ratio,
and the \twCO\ secondary peaks around $81\km\ps$ as complicated broadened features,
suggests that SNR \snr\ is associated
with the molecular gas at a systemic velocity of $\sim81\km\ps$.
There is a discrepancy of $\sim5\km\ps$ between the LSR velocity
of the OH maser and the systemic velocity of the associated MCs.
A similar discrepancy is also seen in SNR~W28, in which the velocity
offset between the OH masers and the main-body MC is as high as
$8\km\ps$ (Claussen et al.\ 1997).

The column density of the molecular gas, $\NHH$, at the OH maser point
(within the beam size $56''$ of the PMOD observation)
is estimated with two methods and presented in Table~\ref{T:mass}.
In the first method, $\NHH$ is calculated by using the \twCO~(\Jotz)
velocity-integrated brightness temperature together with the mean
CO-to-H$_2$ mass conversion factor
$\NHH/W(^{12}{\rm CO})\approx1.8\E{20}~\cm^{-2}\K^{-1}\km^{-1}{\rm s}$
(Dame et al.\ 2001). In the second method, on the assumption
of local thermodynamical equilibrium (LTE) with an excitation
temperature 15~K and \twCO~(\Jotz) being optically thick,
the \thCO\ column density is converted to $\NHH$ using the
relation $\NHH\approx7\E{5}~N(^{13}{\rm CO})$ (Frerking et al.\ 1982).
The results obtained from the two methods are similar.


\subsubsection{The Western Molecular Arc}\label{S:arc}

A close-up tri-color image of the western molecular arc is presented
in Figure~\ref{f:closeup}(a).
As previously mentioned, an X-ray patch is coincident with ``clump 1"
along the arc. The X-ray patch seems to extend northwards
to another small clump (``clump 2"). (The two clumps are labeled in Figure~\ref{f:COmap}.)
A $24\um$ mid-IR patch also seems coincident with ``clump 1".
Another large, bright mid-IR patch in the north part of the image appears
to cover the northernmost CO clump and, projectively,
an ultracompact HII region at the velocity range of 7--$26\km\ps$
(also see Section \ref{S:HII}).
Interestingly, some points with elevated \twCO~\Jtto/\Jotz\
ratios ($>0.9$) at 76--$79\km\ps$ are seen at and around ``clump 1"
(see Figure~\ref{f:COmap}, bottom right panel).
Such ratios significantly higher than the average (0.4--0.6),
as mentioned in Section \ref{S:vlsr}, are also
a signature of warm shocked molecular gas.

For the cores of two of the molecular clumps, two grids of CO spectra are
presented in Figures~\ref{f:clump1} and~\ref{f:clump2},
in which broad \twCO\ line profiles at around $81\km\ps$ can be found.
For ``clump 1", blueshifted (left) broadenings of the \twCO\-line profiles
are seen at points ``1a" (extending to at least $73\km\ps$), ``1b",
``1e", and ``1h". Bumps in the red wings of \twCO\ lines are seen
at ``1b", ``1c", and ``1f" without corresponding \thCO\ emission.
The broad red wing of \twCO\ (\Jotz) at ``1e" is not decided
because the \thCO\ emission is not ignorable ($\ga3\sigma$)
in the 86--$90\km\ps$ interval.
For ``clump 2", blueward broadening of the \twCO\ line wings is seen
at  ``2a", ``2c", and ``2f", while a redward broadening
is clearly seen at ``2d". These \twCO\-line broadenings
in red or/and blue side(s) are as large as about $10\km\ps$,
which is strong evidence that the two clumps are shocked.



\subsubsection{Other Molecular Components along the Line of Sight}
\label{S:other}

\begin{figure*}[tbh!]
\centerline{ {\hfil\hfil
\psfig{figure=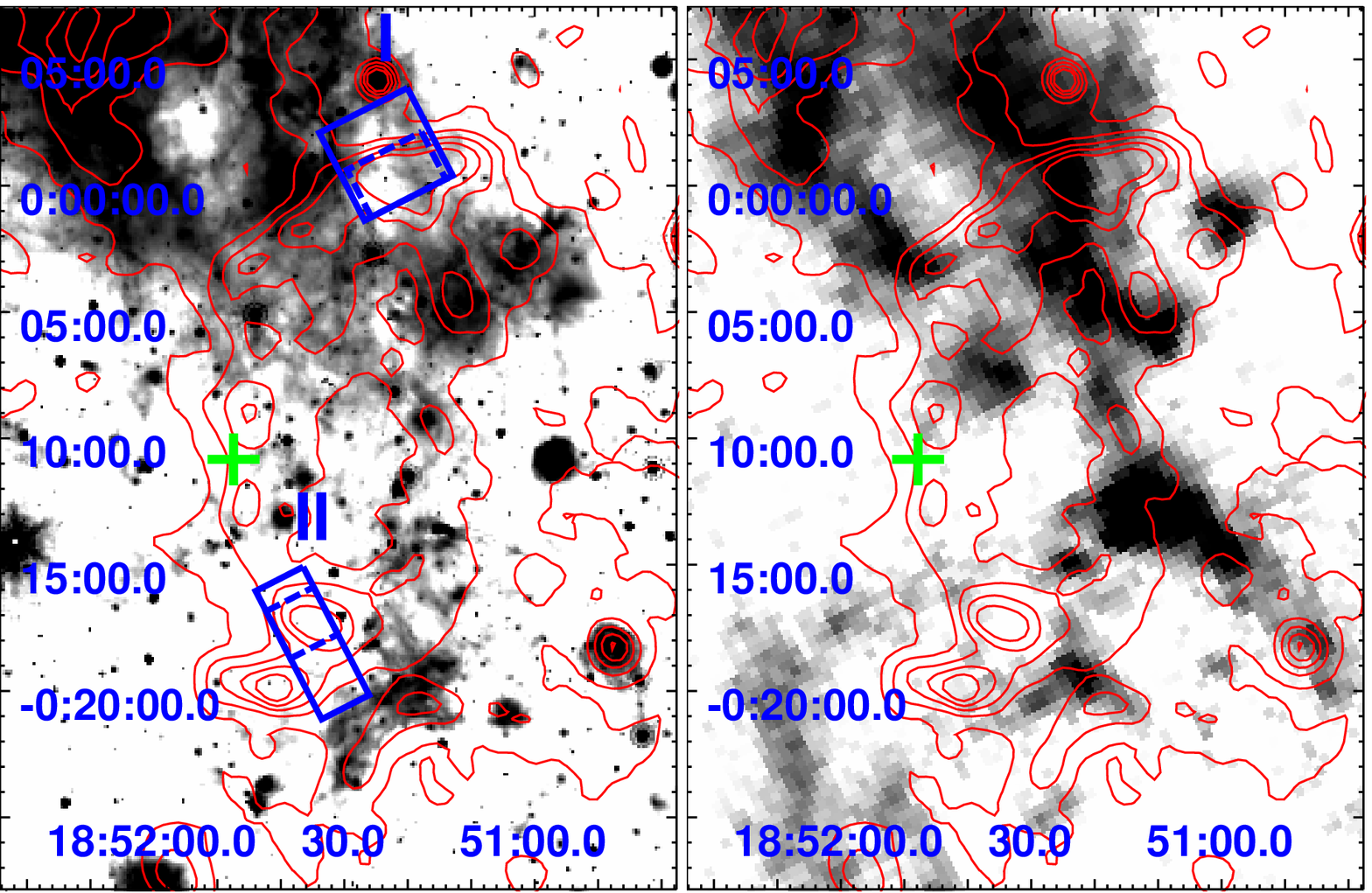,height=2.7in,angle=0, clip=}
\hfil\hfil}}
\caption{
\Spitzer\ $24~\um$ map (left panel) and BU-FCRAO~GRS \thCO~(\Jotz) antenna
temperature map integrated from 99~\kms to 101~\kms (right panel
). Both of the maps are overlaid with 1.4 GHz radio emission contours
with the same levels as those in Figure~\ref{f:COmap}.
The plus sign in each panel denotes the OH maser point.
The two boxes labeled with ``I'' and ``II'' are regions from which
the HI spectra (Figure~\ref{f:HI}) are extracted for discriminating two
candidate distances (Section \ref{S:distance}).
}
\label{f:ridge}
\end{figure*}

\begin{figure}[tbh!]
\centerline{ {\hfil\hfil
\psfig{figure=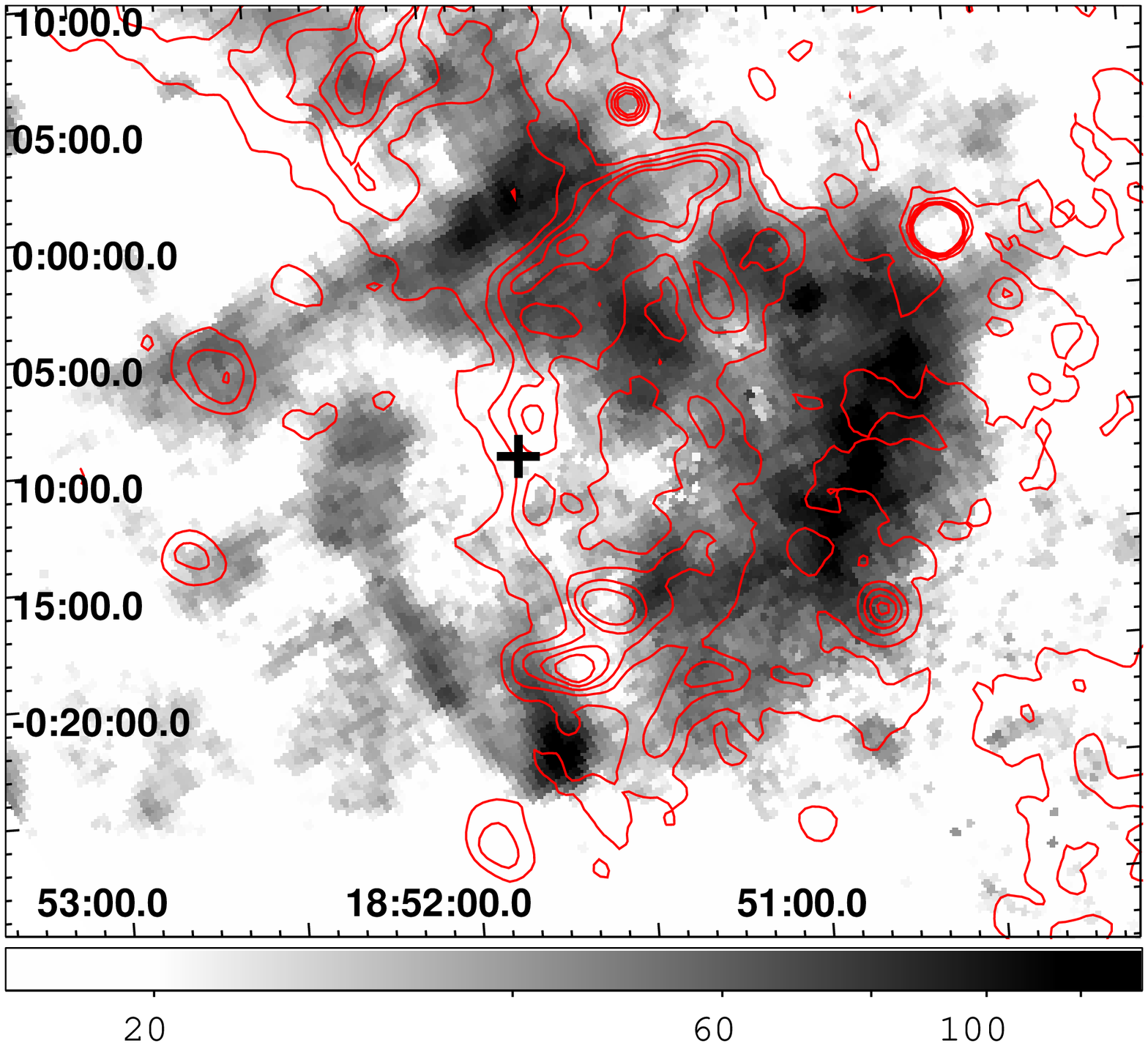,height=3.in,angle=0, clip=}
\hfil\hfil}}
\caption{
Image of the \thCO\ (\Jotz) bubble-like structure at 86--$97\km\ps$
discerned from the BU-FCRAO~GRS data, overlaid with 1.4 GHz radio
emission contours with the same levels as those in Figure~\ref{f:environ}.
The plus sign denotes the OH maser point.
}
\label{f:bubble}
\end{figure}

We also find some noteworthy molecular structures at different
LSR velocities in the \snr\ field.
Figure~\ref{f:ridge} shows the $\VLSR\sim100\km\ps$ \thCO\ (\Jotz) map
compared with the \Spitzer\ $24\um$ mid-IR emission image.
The CO emission at this LSR velocity (the velocity at the tangent point)
is strong in the northeast region of the map (projectively across
the northeastern border of the SNR) and, interestingly, is similar
to the distribution of the bright $24\um$ diffuse emission
(also note a hollow, in the both wavelengths, outsides the remnant).
In both of the \thCO\ and $24\um$ emission maps, there is a
northsouth oriented branch stretching from the northeastern
bright portion and well following the radio continuum
``ridge" that is usually thought to be a structure of the remnant.
At this LSR velocity, however, no molecular features other than the ``branch''
correspond to the other structures of the SNR (e.g., especially,
the eastern radio shell).
As will be discussed in Section \ref{S:ridge}, the faint radio ridge
is likely to be irrelevant to the SNR.

Serendipitously, in the LSR velocity interval 86--$97\km\ps$,
we discern a bubble/ring-like \thCO\ (\Jotz) structure of a radius
$r_b\sim15'$ (see Figure~\ref{f:bubble}) centered at about
($\RA{18}{51}{30}$, $\decl{-00}{09}{00}$, J2000).
Although it appears in almost the same direction as
SNR \snr, with am angular size similar to that of the SNR,
they are not completely coincident with one another
and there is no corresponding feature between them.
The systemic LSR velocity $91\km\ps$ indicates that the bubble/ring
is located at the near distance $\sim5.5\kpc$ or the far distance
$\sim7.9\kpc$, in the background of the SNR
 (see Section \ref{S:distance} for the method for
deriving the kinematic distance).
Using the second method for estimating the molecular column density
described in Section \ref{S:vlsr}),
the mass of the bubble-like molecular gas is estimated to be
$\sim~3.1\E{5}~\Msun$ and $\sim~6.4\E{5}~\Msun$ for the near and far
distances, respectively.
The nature of the structure will be discussed in Section \ref{S:mol@91}.

\subsection{The kinematic distance to SNR \snr}\label{S:distance}

Due to the association of \snr\ with the MCs at the systemic LSR velocity
$\sim81\km\ps$ (Section \ref{S:vlsr}),
there are two candidate kinematic distances to the SNR,
4.8~kpc (near side) and 8.6~kpc (far side).
Here we have used the Clemens' (1985) rotation curve of the Milky Way
together with $R_0=8.0\kpc$ (Reid 1993) and $V_0=220\km\ps$.
The LSR velocity of the OH maser, 86\kms, corresponds to similar
distances 5.2~kpc and 8.3~kpc.

\begin{figure}[tbh!]
\centerline{ {\hfil\hfil
\psfig{figure=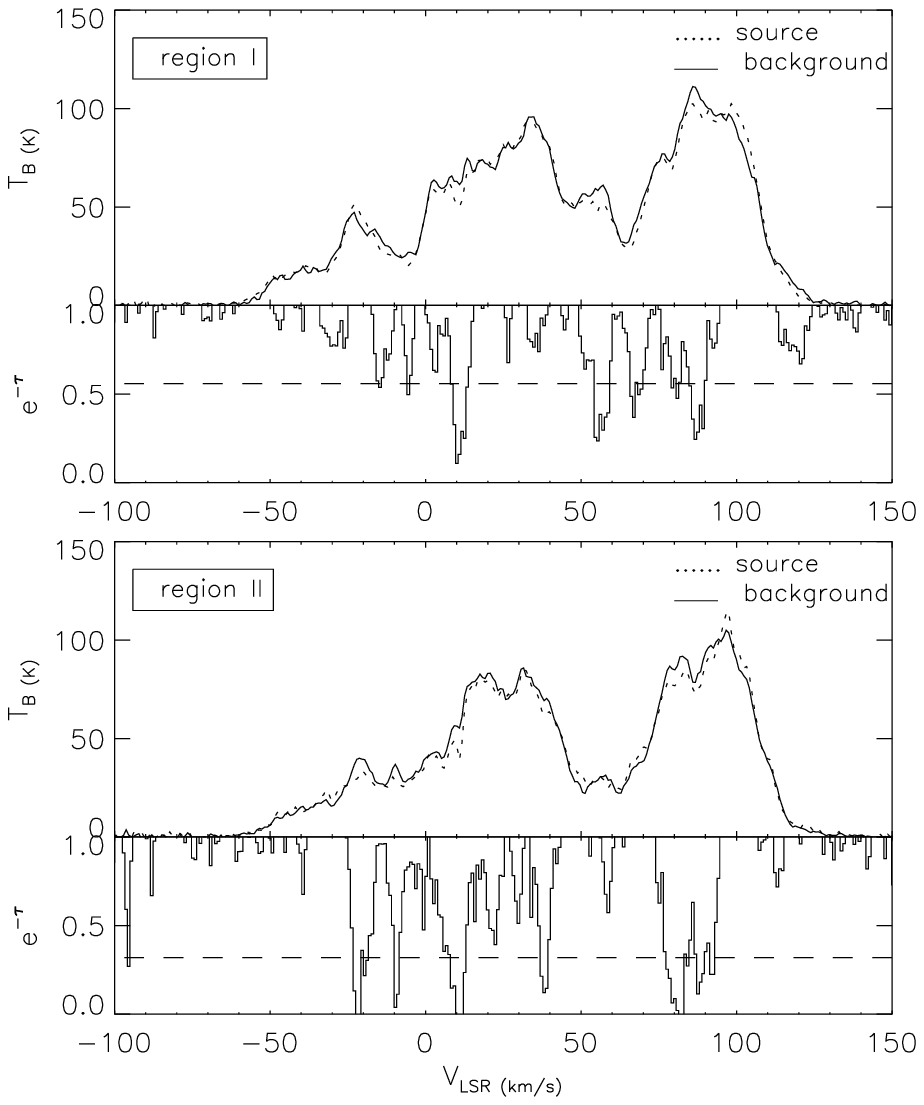,height=4.in,angle=0, clip=}
\hfil\hfil}}
\caption{
VGPS HI spectra extracted from the regions ``I'' and ``II.''
The long-dashed lines represent the $3\sigma$ level.
}
\label{f:HI}
\end{figure}

A modified HI absorption method (Tian et al.\ 2007) is next
used to discriminate between the near and far distances.
The HI spectra are extracted from two radio-bright regions as labeled in
Figure~\ref{f:ridge} (regions ``I" and ``II")
and plotted as shown in Figure~\ref{f:HI}.
A few distinct ($>3\sigma$) absorption features appear
below $\VLSR\sim100\km\ps$,
such as at $10\km\ps$ and in the 80--$90\km\ps$ interval.
However, there is no absorption between 95\kms and 110\kms,
which corresponds to the tangent point (at a distance around 6.7~kpc).
Therefore, SNR \snr\ is in front of the tangent point, namely at a distance
smaller than $6.7\kpc$; hence, the remnant is at the near distance $4.8\kpc$.
Hereafter we parameterize the distance as $d=5\du\kpc$
(also considering the LSR velocity discrepancy of the OH maser
from the systemic velocity).

\begin{figure}[tbh!]
\centerline{ {\hfil\hfil
\psfig{figure=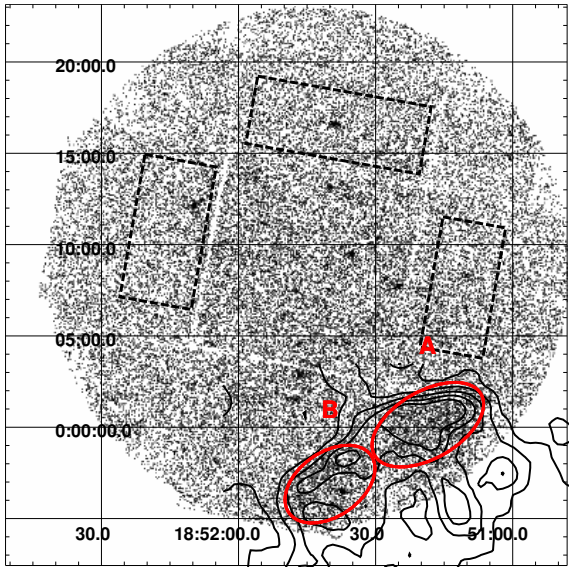,height=2.95in,angle=0, clip=}
\hfil\hfil}}
\caption{
\XMM-\Newton\ raw image of \snr. Two solid ellipses (labeled
``A'' and ``B'') and three dotted boxes are defined, respectively,
for source and background spectrum extraction. The contours of 1.4GHz
radio emission outline the northeastern area of \snr, which have
the same levels as those in Figure~\ref{f:environ}.
}
\label{f:counts_img}
\end{figure}

\begin{figure}[tbh!]
\centerline{ {\hfil\hfil
\psfig{figure=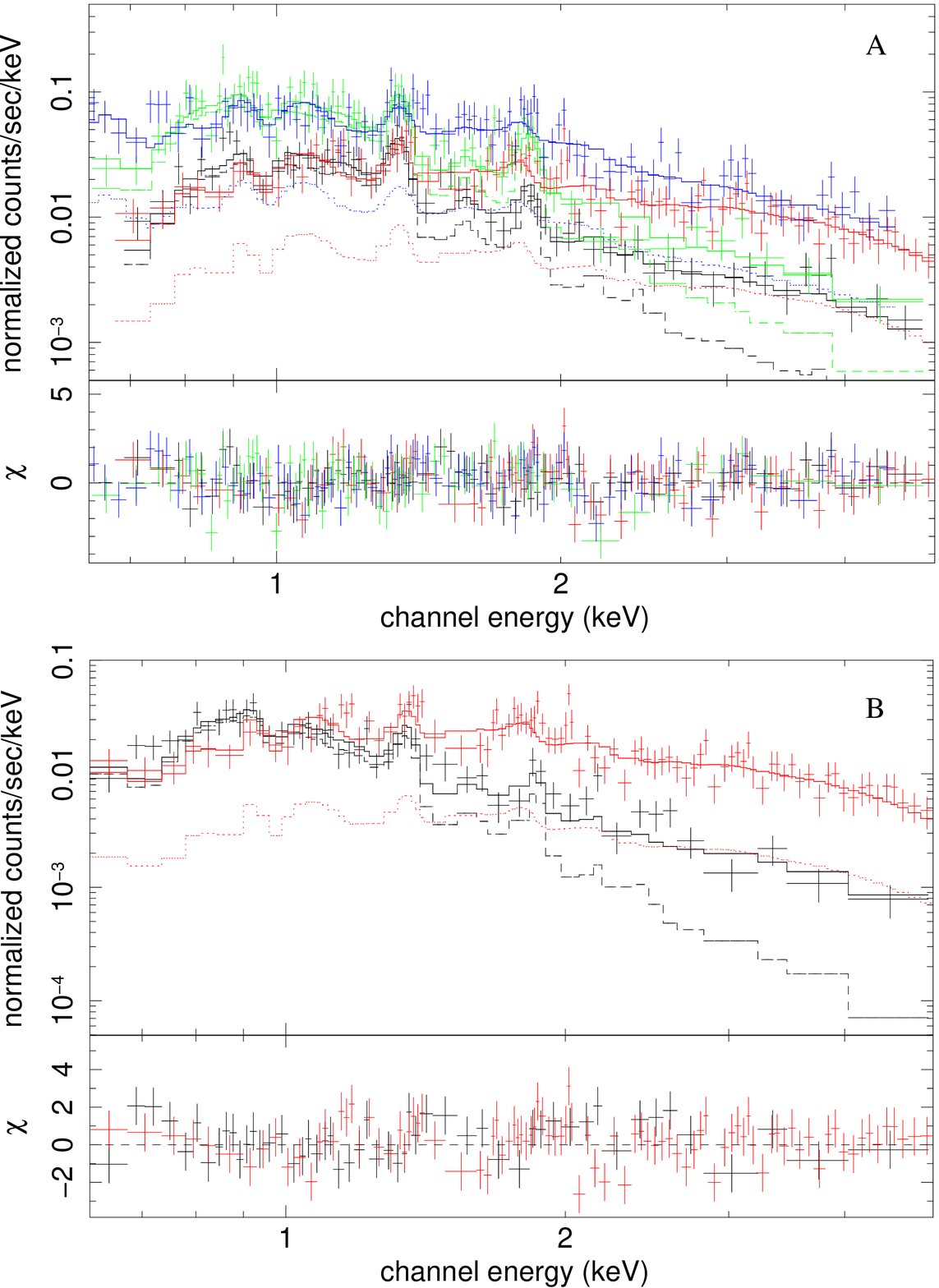,height=4.45in,angle=0, clip=}
\hfil\hfil}}
\caption{
\XMM-\Newton\ EPIC spectra of regions A and B defined
in Figure~\ref{f:counts_img}. The on-SNR spectra (black solid lines
for EPIC-MOS and green solid line for EPIC-PN) are fitted as a
sum of the diffuse background emission and the net SNR emission.
The on-SNR diffuse backgrounds (red dotted lines for MOS and
blue dotted line for PN) are scaled from the off-SNR spectra
(red solid lines for MOS and blue solid line for PN),
which are mimicked by an absorbed {\em nei} + {\em blackbody} model.
The net SNR spectra (black dashed lines for MOS and
green dashed line for PN) are fitted well with
an absorbed {\em nei} model.
}
\label{f:xspec}
\end{figure}

\subsection{\XMM-Newton X-Rays from the Northeastern Boundary}

\subsubsection{X-Ray Image}\label{S:image}
From the available \XMM-\Newton\ EPIC-MOS X-ray data, we produce a tri-color
X-ray image of the northeastern boundary of \snr\ (as shown in
Figure~\ref{f:closeup}, panel (b); 0.5--1.0~keV coded in red,
1.0--2.0~keV in green, and 2.0--7.0~keV in blue),
which is spatially correspondent to the radio brightest region.
In the image production, the background subtraction and exposure
correction have been applied, and adaptive smoothing has been
made with a signal-to-noise ratio (S/N) ratio $>3$.

In the image, the X-ray-emitting region is well bordered by the radio
contours, which indicate a sharp radio brightness drop at the blast shock.
This part of X-ray emission was not detected in the \ROSAT\ X-ray survey.
The eastern portion of the shown region (dominated by the orange
color) is apparently softer than the western portion (dominated by the
green color), which is coincident with the radio brightness peak.


\subsubsection{Spectral Analysis}\label{S:spec}


Before the spectrum extraction, point-like X-ray sources in the FOV were
detected with the wavelet method and removed from the events file.
We then extracted the on-SNR X-ray spectra from two regions (labeled with
``A'' and ``B'', as shown in Figure~\ref{f:counts_img}, the raw MOS
image), which cover the hard (green) and soft (orange)
portions (mentioned above), respectively.
Because of the small amount of available counts,
the MOS1 and MOS2 spectra were merged to increase the statistical quality.
From the EPIC-PN observation, only the spectrum of ``region A" was obtained.
Thus for region A, both the MOS and PN spectra were used simultaneously,
while for region B, only the MOS spectrum was used.
As the X-ray source region is located near the edge of the FOV
and is not very bright relative to the instrumental background,
a double background subtraction method was used.
The off-SNR source-free spectra were extracted from both the MOS and PN
data (the extraction regions, as also shown in Figure~\ref{f:counts_img},
do not include the significant X-ray emission from the Galactic ridge
in the center of the FOV).
The respective instrumental background contributions from the same
on- and off-SNR regions were obtained from the Filter Wheel Closed
data of MOS and PN and were subtracted from the on- and off-SNR
spectra.
The individual on- and off-SNR spectra were adaptively binned to
achieve a background-subtracted S/N of 3.

\begin{center}
\begin{deluxetable*}{l|cc}
\tabletypesize{\footnotesize}
\tablecaption{Spectral Fitting Results for the Northern Shell of Kes 78 with 90\% Confidence Ranges and Estimates of the Gas Density}
\tablewidth{0pt}
\tablehead{
\colhead{Regions} \vline & \colhead{A}  & \colhead{B}
}
\startdata
Net count rate ($10^{-2}$ counts$\ps$) & $3.61\pm0.11$ (MOS) & $2.79\pm0.10$ (MOS)\\
    & $8.52\pm0.27$ (PN) & \\
$\chi_{\nu}^{2}$ (dof) & 1.04 (318) & 1.17 (137)\\
$\NH$ ($10^{22}\cm^{-2}$) & $1.04^{+0.09}_{-0.08}$& $0.70^{+0.12}_{-0.09}$\\
$kT_x$ ($\keV$) & $1.51^{+0.34}_{-0.50}$& $1.30^{+1.57}_{-0.65}$\\
$n_e t_i$ ($10^{10}\cm^{-3}\,{\rm s}$) & $1.82^{+0.89}_{-0.32}$& $2.03^{+4.88}_{-0.72}$ \\
$f\nel\nH V/\du^{2}$ ($10^{56}\cm^{-3}$) & $1.12^{+0.31}_{-0.25}$& $0.58^{+0.69}_{-0.29}$\\
Flux($10^{-12}\erg\cm^{-2}\ps$)$^{\rm a}$ & $3.41^{+0.94}_{-0.77}$ & $1.82^{+2.19}_{-0.92}$\\
$\nH/f^{-1/2}\du^{-1/2}$ ($\cm^{-3}$)$^{\rm b}$
   & $0.11^{+0.01}_{-0.01}$& $0.10^{+0.05}_{-0.02}$
\tablecomments{
   The net count rates listed in the table are for the on-SNR spectra.
   The rates of the off-SNR spectra are $6.17\pm0.18\E{-2}$ counts$\ps$
   and $1.26\pm0.04\E{-1}$ counts$\ps$ for MOS and PN, respectively.
}
\enddata
  \tablenotetext{a}{\phantom{0} The unabsorbed fluxes are
   in the 0.6--$5.0\keV$ band.}
  \tablenotetext{b}{\phantom{0} In the estimate of the densities, we assume
    oblate spheroids for elliptical regions A (with half-axes
    $3.'36\times3.'36\times1.'84$) and B ($2.'74\times2.'74\times1.'75$).
}
\label{T:xpara}
\end{deluxetable*}
\end{center}

The XSPEC spectral fitting package (ver.11.3)
\footnote{http://heasarc.gsfc.nasa.gov/docs/software/lheasoft/xanadu/\\
xspec/xspec11/index.html}
was used.
Each on-SNR spectrum was jointly fitted together with the off-SNR
spectrum (Figure~\ref{f:xspec}).
The on-SNR diffuse background was determined by scaling the off-SNR
emission according to the region sizes (Figure~\ref{f:xspec}) and is
phenomenologically well described by an absorbed {\em nei} +
{\em blackbody} model.
For the foreground absorption, the cross sections from Morrison
\& McCammon (1983) were used, and solar abundances were assumed.
The spectra of both regions A and B show distinct line features
of Mg~He$\alpha$ ($\sim1.35\keV$) and Si~He$\alpha$ ($\sim1.85\keV$),
indicating the thermal origin of the emission.
The net SNR X-ray emission can be well fitted with an absorbed,
non-equilibrium ionization thermal plasma model ({\em vnei})
with solar abundances.
The fitting results are summarized in Table~\ref{T:xpara}.
The physical properties between the two regions are similar,
except that the intervening hydrogen column density for region~A
is higher than that for region~B. The relatively high $\NH$ for
region A is consistent with a contact of the SNR at the
northeastern border (coincident with region~A) with the dense
molecular strip (the shock strip interaction can naturally explain
the radio brightness peak there;
see Section \ref{S:vlsr} and Figure~\ref{f:environ})
and explains the fact that X-rays from region~A are harder than
those from region~B (see Section \ref{S:image}).
The gas temperatures are found to be 1.0--$1.9\keV$ for region~A
and 0.7--$2.9\keV$ for region~B.
The ionization timescales $n_e t_i$ of the X-ray-emitting gas
of the two regions are both on the order of $10^{10}\cm^{-3}{\rm s}$,
much smaller than $10^{12}\cm^{-3}{\rm s}$, which
indicates that the hot gas has not yet reached ionization equilibrium.
Estimates of the hydrogen number densities, $\nH$,
derived from the volume emission measures
($fn_e\nH V$, where $f$ is the filling factor of the hot gas
and $n_e\sim1.2\nH$ is assumed) are also given in Table~\ref{T:xpara}.
In the derivation of the densities we have assumed that the
three-dimensional shapes of the elliptical regions (A and B) are ablate
spheroids.
Deviations from these assumptions and the non-uniformity of
the X-ray-emitting plasma are consolidated into
factor $f$ of the individual regions.

\section{Discussion}

\subsection{Global Evolution of the SNR and the Progenitor} \label{S:evolution}

As revealed above (Section \ref{S:vlsr}), SNR~\snr\ evolves in an interstellar
environment of molecular gas with systemic velocity $\sim81\km\ps$
and expands into a cloudy region in the east.
Interestingly, the X-ray-emitting gas in the northeastern edge
is characterized by a single thermal temperature (around 1.5~keV)
and a low number density ($\nH\sim0.1f^{-1/2}\du^{-1/2}\cm^{-3}$).
This indicates that the blast wave in the northeast of the SNR
propagates in the intercloud/interclump medium (ICM).
%
%
%


The velocity of the blast wave propagating in the ICM can be estimated
as $v_s=[16kT_x/(3\bar{\mu}\mH)]^{1/2}\sim1.1_{-0.2}^{+0.1}\E{3}\km\ps$,
where the fitted hot gas temperature for region A (for which the error
bars of the spectrally fitted parameters are smaller than those for
region B), $kT_x=1.51^{+0.34}_{-0.50}\keV$, is adopted as the postshock
gas temperature and the mean atomic weight $\bar{\mu}$ equals 0.61 for
fully ionized plasma.
We take the curvature radius of the eastern radio shell $\sim12'$
as the SNR radius, namely, $\rs\sim17\du$~pc, assuming that the dynamical
age of the remnant, $t\sim 2\rs/(5\vs)\sim 6.1_{-0.7}^{+1.0}$~kyr,
is derived with the Sedov (1959) evolution law.
The explosion energy is
$E=(25/4\xi)(1.4n_0\mH)\rs^3\vs^2\sim4.7\E{50}f^{-1/2}\du^{5/2}\erg$,
where $\xi=2.026$ and $n_0=n_{\rm H}/4$ is the
preshock intercloud hydrogen density.
The ionization age of the X-ray emitting gas in the northeastern
edge is inferred from the ionization timescales (see Table~\ref{T:xpara});
for region A, $t_i\sim4.4_{-0.9}^{+2.2}$~kyr old,
which is slightly, but reasonably, smaller than the dynamical age.
Both estimates suggest a relatively young age of \snr.

The absence of [OIII] leads to an inference that the velocity
of the shock wave responsible for the optical filamentary and diffuse
structures is $<100\km\ps$ (Boumis et al.\ 2009).
This shock cannot be the SNR blast wave, otherwise the SNR
should have been in the radiative phase
with a considerably large age (Chevalier 1974): $0.31\rs/\vs\ga52$~kyr,
inconsistent with the X-ray properties obtained above.
As Boumis et al.\ (2009) suggest, however, it is in the
interstellar clouds that the slow shocks are propagating.

The bright radio emission in the east of the SNR is in agreement
with the aggregation of the $\sim81\km\ps$ molecular gas there
and the inference that the clouds are clumpy.
The radio continuum emission may arise from the blast shock
that propagates in the ICM (Blandford \& Cowie 1982).
Contrarily, in the west, the $\sim81\km\ps$ CO emission is weak
and the radio continuum fades out.

As also mentioned in Section \ref{S:vlsr}, 
in the west, the SNR expands into a relatively low-density region;
but there the SNR seems to collide with a clumpy molecular arc.
The entire SNR appears to be in a cavity surrounded by
molecular matter.
The molecular number density of ``clump 1" along the arc is
$\sim270\du^{-1}\cm^{-3}$/$188\du^{-1}\cm^{-3}$,
with an angular radius $2'$ adopted.
The broad CO line wings and enhanced \twCO~\Jtto/\Jotz\
ratios along the western arc impliy a disturbance suffered
by the molecular gas.
Similar molecular arcs are seen in some other interacting
SNRs, such as Kes~69 (Zhou et al.\ 2009) and Kes~75
(Su et al.\ 2009), in which the arcs are suggested to represent
the debris of the clumpy shells of the progenitors' wind bubbles.
If the molecular cavity in \snr\ was created by the progenitor,
using the linear relation between the main-sequence wind bubble
size and the progenitor's mass (Y. Chen, \etal.\ 2011, in preparation
\footnote{http://astronomy.nju.edu.cn/$\sim $ygchen/papers/Rb-M/Rb-M.pdf}),
$\Rb=1.21M/\Msun-8.98$~pc,
the cavity radius 17pc would imply a mass around $22\Msun$ for the
progenitor star of SNR \snr, which seems to be of O9 type.



\subsection{Pressures in the multi-phase gases}
We have arrived at such a scenario in Section \ref{S:evolution}
that SNR \snr\ evolves in a cloudy interstellar medium,
the blast wave in the ICM giving rise to the X-rays
and the cloud shocks in the dense clouds to the optical emission.
The thermal pressure of the X-ray-emitting gas
is $2.3\nH T_x\sim6.0\E{6}f^{-1/2}\du^{-1/2}\cm^{-3}\rm K$ and
$\sim4.0\E{6}f^{-1/2}\du^{-1/2}\cm^{-3} \rm K$
for regions A and B, respectively.
According to the density-sensitive line ratio of
[SII]$\lambda\lambda6716/6731$, B09 determined the electron densities
in the shocked clouds to lie below $200\cm^{-3}$ and suggest a
temperature of the electrons $\sim10^4$~K.
The upper limit of the thermal pressure in the optically emitting
structures is estimated to be $\sim4\times 10^6 \cm^{-3} \rm K$,
which is similar to the pressure derived for the X-ray-emitting gas.
Thus, it seems that the shocked ICM may be in pressure balance with
the shocked clouds.

However, there is significant pressure variation
in the multi-phase cloudy gases in the east of the remnant.
The 1720~MHz OH masers are believed to arise from very dense
($\sim10^5\cm^{-3}$) regions of MCs that are shocked by
the C-type shock waves (Lockett et al.\ 1999).
By means of Zeeman splitting, the LOS component of
the magnetic field in the Kes~78 OH maser region,
$B_{\rm los}$, was measured to be $1.5\pm 0.3 \rm mG$
(Koralesky et al.\ 1998).
For a randomly oriented field, the post shock field
strength $B_{\rm ps}=2B_{\rm los}$ (Frail \& Mitchell 1998;
Crutcher 1999).
Thus, the magnetic pressure in the OH maser region of \snr\
is $B_{\rm ps}^2/(8\pi k)\approx2.6\E{9}\cm^{-3} \rm K$,
which is over two orders of magnitudes higher than the thermal
pressures of the X-ray emitting intercloud plasma and the
optically emitting filamentary and diffuse structures.
Notably, similar pressure variation from the X-ray emitting gas
to the OH maser regions is seen in SNR~W28 (Rho \& Borkowski 2002).
Such a pressure contrast has been suggested to be caused by
the interaction of the radiative shell with the dense molecular
clumps (Chevalier 1999).

\subsection{The Dubious Ridge} \label{S:ridge}
The radio ``ridge" was sometimes suspected to be the western part of an
elongated SNR shell (e.g., Caswell et al.\ 1975).
We have seen in Section \ref{S:other} that the radio ridge appears
to be coincident with the northsouth oriented $24\um$ IR branch,
which extends from the bright emission in the northeastern region
outside the remnant's radio boundary, the $24\um$ emission in and
around the eastern region of the SNR has a strikingly similar
distribution to the $\VLSR\sim100\km\ps$ \thCO\ emission.
This leads us to suspect that the radio ridge is not a part of
the SNR that is associated with the $\sim81\km\ps$ MCs, but
rather is associated with the $\sim100\km\ps$ MCs,
which is essentially located at the tangent point of the inner
Scutum--Centaurus spiral arm of the Galaxy.
On the other hand, optical observation towards \snr\ finds
no significant [SII] from the radio ridge,
while in the eastern, northern, and southern regions of \snr,
the [SII] emission is evident, with the [SII]/\Ha\ ratio $>1.3$
(Boumis \etal\ 2009). As the [SII] strength in the shock-ionized gas
is comparable to the \Ha\ strength and is much fainter in the
photoionized gas (HII regions or planetary nebula;
Dennefeld \& Kunth 1981), the radio emission is probably of
thermal origin and possibly arises from star formation region(s).
Furthur study estimating the radio spectral index for the
radio ``ridge" will test its nature.

\subsection{Other Sources in the Kes 78 Field}
\subsubsection{PSR1850-0006}
Since the progenitor of \snr\ was probably a massive star
(see Section \ref{S:evolution}),
a neutron star or pulsar may have been left after its explosion.
PSR1850-0006 is projected in the northwestern portion
of the SNR (Keith et al.\ 2009).
However, with a rotational period of 2.2~s and a period derivative
of $4.3~\E{-15}$~s$\ps$,  the pulsar has a characteristic age of
about 8~Myr, which is much older than the age of SNR \snr\
(Section \ref{S:evolution}), disfavoring an association between
PSR1850-0006 and the SNR.

\subsubsection{HII Regions} \label{S:HII}
The ultracompact HII region located at ($\RA{18}{51}{25}$,
$\decl{00}{04}{07}$, J2000)
is northeast of the SNR, as seen in Figure~\ref{f:environ}.
It is coincident with the massive star formation region IRAS~18488+0000
(Bronfman \etal\ 1996) and may be associated with SNR \snr.
In fact, a 22~GHz ${\rm H_2O}$ maser at $\VLSR\sim~80\km\ps$, which is very
similar to the systemic LSR velocity of the MCs with which
the SNR is in contact, is suggested to be associated with the
star formation region and is at the same distance (5~kpc)
as we determine here for the SNR (Beuther et al.\ 2002).
The HII region or IRAS~18488+0000
is located at the northern molecular strip with which the SNR interacts
and only $2'$ north of the radio brightness peak of the SNR.

Another ultracompact HII region at ($\RA{18}{50}{31}$,
$\decl{-00}{01}{55}$, J2000) (Garay et al.\ 1993) appears to be
projectionally located on the northwestern boundary of the SNR
and connects the western molecular arc in the northwest, as
seen in Figure~\ref{f:environ}.
In this HII region, six 22~GHz ${\rm H_2O}$ masers were detected
at the LSR velocity range $\sim7$--$26\km\ps$ (Hofner \& Churchwell 1996).
There may not be associated with the SNR because they are at different
LSR velocities.

\subsubsection{The Molecular Structure at $\VLSR\sim91\km\ps$}
\label{S:mol@91}
As concluded in Section \ref{S:other}, the bubble/ring-like structure
discovered at $\VLSR=86$--$97\km\ps$ is at a furthur distance
(5.5/7.9~kpc) than that
to SNR \snr\ (4.8~kpc).
If it is an interstellar bubble, then the mean molecular number density
of the undisturbed gas before the bubble was created
is $\sim77/54\cm^{-3}$ for the near/far distance.
Adopting an expansion velocity $v_b\sim7\km\ps$ (FWHM of the \thCO\ line
at $\sim91\km\ps$),
the kinematic energy is $\sim1.5/3.1\E{50}\erg$.
If the bubble is blown by a wind , it has an age
$3r_b/(5v_b)\sim2.0/2.9\E{6}\yr$ and the
mechanical luminosity of the wind is $\sim1.2/1.8\E{37}\erg\ps$ according
to the canonical evolutionary law of a wind bubble given by Weaver
et al.\ (1977).
This ring/bubble-like structure may not be a distinct SNR.
Actually, for an SNR in the radiative stage (Chevalier 1974), its age is
$0.31r_b/v_b\sim1.0/1.5\E{6}\yr$, but the supernova explosion energy
would be $4.6/9.6\E{51}\erg$, which is much higher than the fiducial value
$1\E{51}\erg$.

\section{Summary}
We have investigated the molecular environment of the Galactic
SNR~\snr\ using the CO observations in PMOD, KOSMA, and FCRAO
and have performed an \XMM-\Newton\ X-ray spectroscopic study
for the northeastern edge of the remnant.
The main results and conclusions are summarized as follows:

\begin{enumerate}
\item SNR~\snr\ is found to be associated with the MCs at a systemic
LSR velocity of $81\km\ps$.
At around this velocity, the SNR is revealed by the \thCO\
observation to be in the side of some dense MCs
and is consistent in extent with a cavity of \twCO\ gas.
The SNR expands into a cloudy dense region in the east,
where the OH maser emission arises and the bright radio shell overlaps with
most of the region of strong \twCO\ emission.
Broadened \twCO\ line profiles discerned in the eastern maser region
and the western clumpy molecular arc,
the elevated \twCO~\Jtto/\Jotz\ ratios along the SNR boundary
 may be signatures of shock perturbation in the molecular gas.

\item The association of SNR \snr\ with the $\sim81\km\ps$ MC,
together with the HI absorption along the LOS, places the SNR
at a kinematic distance of 4.8~kpc.


\item The X-rays arising from the northeastern radio shell are
emitted by underionized hot ($\sim1.5\keV$), low-density
($\sim0.1\cm^{-3}$) plasma with solar abundance,
and the plasma may be of intercloud origin.
Especially, the X-rays from the radio brightness peak, where a long
molecular strip appears to connect, suffers a heavier absorption than
those from other parts along this \XMM-observed section of the shell.
The thermal pressures of the X-ray-emitting gas and the optically
emitting structures along the shell are lower than the magnetic
pressure in the OH maser region by up to two orders of magnitudes.
The age of the remnant is inferred to be about 6~kyr.

\item The size of the molecular cavity in \snr, which is assumed to be
created by the main-sequence wind, implies the progenitor's
mass to be around $22\Msun$.

\item The north-south oriented radio continuum ridge west of
the bright shell is suggested to be irrelevant with the SNR,
but related to a $\sim100\km\ps$ CO branch at the tangent point,
which corresponds to IR bright, low-[SII]/\Ha\ ratio
star formation region(s).

\item Along the LOS toward SNR~\snr, we discern an irrelevant
bubble/ring-like \thCO\ structure $15'$ in angular radius at the LSR
velocity $91\km\ps$, but the possibility of it being a separated SNR is
ruled out.

\end{enumerate}

\begin{acknowledgements}
The authors are thankful to the staff members of the KOSMA observatory
and Qinghai Radio Observing Station at Delingha for their support in
observation. We also thank Yang Su, Xin Zhou, and Junzhi Wang for helpful
discussion. We acknowledge the use of the VGPS and GRS data; the
National Radio Astronomy Observatory is a facility of the National
Science Foundation operated under cooperative agreement by
Associated Universities, Inc. This work is supported by the NSFC grants
10673003 and 10725312 and the 973 Program grant 2009CB824800.
\end{acknowledgements}


\end{document}